\documentclass[preprints,article,accept,moreauthors,pdftex]{Definitions/mdpi}
\firstpage{1} 
\pubvolume{1}
\issuenum{1}
\articlenumber{0}
\pubyear{2022}
\copyrightyear{2022}
\datereceived{} 
\dateaccepted{} 
\datepublished{} 
\hreflink{https://doi.org/} % If needed use \linebreak
\Title{Surface waves prediction based on long-range acoustic backscattering in a mid-frequency range}

\TitleCitation{Title}

\Author{Alexey V. Ermoshkin$^1$\orcidB{}, Dmitry A. Kosteev$^1$,  Alexander A. Ponomarenko$^{2,1}$*\orcidA,
    Dmitry D. Razumov$^1$*, Mikhail B. Salin$^1$*\orcidC{}}

\address{%
$^{1}$ \quad Center for hydroacoustics and Geophysical research division, Institute of Applied Physics of the Russian Academy of Sciences, Nizhny Novgorod, 603950, Russia\\
$^{2}$ \quad Laboratory of Algorithm and Technologies for Network Analysis, HSE University, Nizhny Novgorod, 603036, Russia}

\corres{Correspondence: aponomarenko@hse.ru (A.P.); ddrazumov@ipfran.ru (D.R.); mikesalin@ipfran.ru (M.S.)}

\abstract{New data was obtained for a frequency band that had not been so well-studied for sea surface probing applications before. During the described 2-weeks sea experiment 1-3 kHz tonal pulses were emitted from a platform, located on the northern Black Sea shelf, and Doppler spectrum of reverberation was studied. We believe that this band is worth further studying due the sound propagation range is large enough to meet practical needs in coastal zone while the angle-distance resolution is quite moderate. However it is quite difficult to interpret the obtained data since backscattering spectrum shape is influenced by a series of effects and has a complicated link to wind waves and currents parameters. Backscattering of acoustical signals was received for distances around 2 nautical miles. Significant wave height, dominant wave frequency  were estimated as the result of such signals processing with the use of machine learning tools. A decision-tree-based mathematical regression model was trained to solve the inverse problem. Wind waves prediction is in a good agreement with direct measurements, made on the platform, and machine learning results allow physical interpretation.}


\begin{document}

% -----------------------------------------------------------------------
% ---== MAIN TEXT ==---

\section{Introduction}
%{\it Mike writes the introduction.}
Monitoring and forecasting of currents and wind waves is necessary for activities such as ship navigation, and routine operation of ports and offshore platforms. Studies in different areas of ocean physics rely on statistics of long-term observations. Different methods and instruments are used to do such measurements and collected data might be used, for example, to validate models and to assimilate data in real-time forecasting \citep{Kuznetsova2016,Wei2022}.

Huge amount of remote measurements tasks in the sea are fulfilled with use of radars and other microwave instruments. The idea of using radars to determine the kinematic characteristics of wind waves is hugely attractive, primarily due to the simplicity of measurements, and such evident advantages of radar sensing as all-weather and day-and-night operation. The methods of extracting the period and direction of the waves from the radar data are well-developed \citep{Young1985}. The possibility of measuring the wave height based on the data on the backscattering intensity has been studied in a number of papers, e.g., in \cite{Dankert2005}. In contrast to the backscattering intensity, the Doppler frequency shift of the reflected radio wave is associated with the radial velocity of the scattering elements. The variable component of the Doppler shift is determined by the orbital velocities of the wind waves. This is the principle underlying operation of a radar device used to determine wave heights \citep{Rosenberg1973}. Studies of Doppler characteristics of microwave radio waves scattering were described in a large number of papers, both theoretical and experimental, starting from the 1960s. Currently, the main model used to explain the features of the Doppler spectra is the two-scale scattering model \citep{BassFuks1972}. The possibility of determining the wave height and surface currents using coherent radar data at a low grazing angle was studied in \cite{Lyzenga2010,Carrasco2017,Ermoshkin2019,Ermoshkin2020}. Those works used the approximation of linear free-surface waves to find a relation between the wave height and its orbital velocity measured with a Doppler radar. It has been shown that the wave spectra and currents reconstructed from velocity radar images correlate with the data of contact measurements, both qualitatively and quantitatively. 

Usage of acoustic measurement tools is attractive too, e.g. bottom-mounted self-contained devices can operate when above-the-water operations are stopped due to weary conditions like heavy storm or presence of drifting ice. Acoustic Doppler current profiler (ADCP) is a very widely-used tool to measure speed in a water column, and modern models are facilitated with a function to measure surface waves spectra as well \citep{Adibzade2021}. ADCP is equipped with 4 echo sounder beams that can be treated as 4 water height gauges than allows tracking elevation and skew of the surface, produced by long waves. Another design of upward-looking acoustic tools relies on sound scattering model to reconstruct surface spectrum moments basing on the Doppler spectrum of the reflected signal \citep{Titchenko2019a,Titchenko2019b,Salin2016}. High frequency signals reverberation patterns may allow to track individual wave crests in vicinity of the transducer, as it has been reported by \cite{Badiey2014} and \cite{Richards2021}. The mentioned tools are designed for local measurements and methods to assess waves and currents on a long range are of great interest too.

Recently \cite{CAT1}, \cite{CAT2} and \cite{CAT3} showed great progress in coastal acoustic tomography. This technique allows estimating mean current in a channel or a strait basing on acoustic pulses travel times between multiple sources and receivers, organized in a net. This is how ocean acoustic tomography technique, introduced in 1980's for making measurements on the scales of hundreds of kilometers, is adapted to the scales up to 5 km nowadays. The frequency band has been raised to about 5 kHz respectively. A group of bottom-mounted transmitter-receivers can also allow to track passage of internal waves through the such field of sensors \citep{Serebryany2013}.

An acoustic environmental monitoring system can be designed analogous to a radar. Underwater sound scattering on surface roughness is very much analogous to microwave scattering. E.g. \cite{Dahl1997} tested simultaneous illumination of a rough surface by microwaves from above and sound from below and showed that if their parameters are matched, acoustic scattering strength match the one for microwaves until bubbles start to appear in large amount. When one takes grazing angles, underwater sound propagation in not limited by direct line of sight due to the waveguide effects. Compact devices can transmit sound pulses of the frequency of several kilohertz, that will travel several kilometers. Long-range measurements are going to be based on averaged interaction of sound with surface waves, so that scattering intensity is accumulated over a large patch of the surface: around ten surface wavelengths in diameter or larger.

Small perturbation theory (SPT) gives a rather straightforward relation between the power spectral density (PSD) of the scattered signal and surface waves PSD due to Bragg resonance of sound waves with surface waves \citep{Salin2012,BJORNO2017}. A measurement method was proposed by \cite{Dolin1995}, directly based on this effect, but the were no report that it had ever been realized. \cite{Hayek1999} presented the experimental data as up-Bragg and down-Bragg levels and showed that the direction of the most intensive scattering signal arrival is correlated with the wind direction. However a spectrum of the scattered signal was not shown it that paper. 

SPT-based models become not enough precise if the sea surface is too rough in comparison with the sound wavelength or the bubbles that are not included into model appear in large amounts. \cite{Bjorno2001} noticed that backscattering level differs from the one, predicted by linear theory, starting from the wind speed 7-10 m/s and frequency of 300-500 Hz. That is called anomalous low frequency surface reverberation. The phenomenon is explained by scattering on cavernas and bubble, which appear due to wave breaking. E.g. echo sounder measurements \citep{Thorpe1982} showed presence of bubble clouds on the depth of 3 m at a wind speed of 6-8 m/s. More recent study \citep{Deane2002} revealed that such clouds may contain a measurable amount of bubbles of 1-3 mm diameter. So bubbles from breaking waves may influence on sound propagation and scattering not only in the high frequency band but also in the band around 1 kHz \citep{Akulichev2019,Liu2021}. For this reason \cite{Salin2018} studied a problem how does the presence of bubble is revealed in scattering of Doppler-sensitive pulses and showed that the bubble scattering has a specific signature in backscattering Doppler spectra. Subsurface bubbles move definitely slower than typical phase speeds of wave crests. They are carried by oscillating wind-waves-driven currents as well as by the mean water currents in upper layer. 

Doppler surface scattering prediction in 1-3 kHz band is often studied the context of sonar performance modeling \citep{Doisy2008, Rui2016, Colin2021}. The present research is devoted to the inverse problem of predicting a sea surface state basing on backscattering signal. What we are going to analyze here is actually the reverberation signal. 

To test the possibility of reconstruction of such parameters a continuous complex experiment was carried out. A sonar, made of researchers equipment and based on a sea platform, was working in the same regime during 14 days of the experiment, while the weather conditions were in different ways. The experiment took place in the shelf zone of the northern part of the Black Sea in the second half of September 2021. All necessary parameters, concerning wind, waves and currents were recorded in parallel with direct measurement tools. It was developed an acoustic signal processing algorithm that included adjustable parameters to reconstruct the sea surface state. First, we tried manually adjust correlations, seeking correlation between input and output parameters and making calibration curves. Second, we found more practical to apply machine learning methods to do this. Extreme Gradient Boosting (XGBoost) method, based on decision trees \citep{chen2016xgboost}, was used for that. To make it applicable, part of the data (random time intervals) was treated as training dataset, and the rest was used for testing. 

%---------------------------
\section{Theory background}
\label{sec:theory}  %{\it This section is for Mike}
To understand the sound scattering phenomenon in this experiment one should keep in mind that the rough surface is insonified not only by the direct arrival, but by the sound waves, propagating by various paths with surface and bottom reflections. Though sophisticated methods of computing reverberation in a shallow water waveguide exist \citep{Ellis1995, Kazak2021}, in practice we may still relay on the sonar equation  for overall estimates \citep{Urick1983, BJORNO2017, LYNCH2017}. The model is the following. A sound wave experience many quasi-mirror reflections i.e. when the surface and the bottom are considered to be flat, preserving their mean levels. Each time a small part of the energy is scattered backward (and this part is too small to be accounted for extra decay of the forward propagating wave). Back propagation of scattered waves is modeled in so that and only quasi-mirror reflections are taken into account. Scattered signals reach the receiver with random phases, so only accumulations of intensities are needed. 

Since all properties, that define propagation, remain stable during the experiment, all changes in scattering levels due to weather conditions should be linked to the scattering strength (SS) parameter \citep{Urick1983}. Chapman and Harris provided an empirical law for SS as a function of a wind speed, a frequency and other parameters \citep{Urick1983, BJORNO2017}. A strong contribution of bottom scattering is expected in shallow water, so a level of Doppler side lobes in the spectrum of the scattering signal (i.e. PSD of SS) is going to be more reliable that total SS as itself. Small perturbation theory might have help with theoretical analysis of such features, but it is valid in case of the Rayleigh condition is satisfied:
\begin{equation}
    P_R = 2 k a_{std} sin(\chi_{\star}) << 1
    \label{eq:mbs_rayligh}
\end{equation}
where $k$ is the acoustic wavenumber, $a_{std}$ is a root mean square of surface elevation, and $\chi_\star \approx 30^{\circ}$ is a waveguide capture angle (that stands here because it is an estimate for the maximum grazing angle of an incident wave). The value of $a_{std}$ is related to a more common oceanography parameter, a significant wave height as $H_{m0} = 4 a_{std}$. Treating 'much less' as 'equivalent with coefficient 0.16' allows to conclude that at the frequency of 1.5 kHz small perturbation theory is valid up to $H_{m0}=10$ cm. Unfortunately, such low amplitude was observed only once during a half-hour session of the whole 2-week experiment. 

Moreover to that the $k-\omega$ spectrum of surface waves is known to diverge from its trivial form like:
\begin{equation}
    G^2(\textbf{K},\Omega) \approx \frac{g^2}{4\pi\Omega^3} S(\Omega) \delta \left( \Omega - \sqrt{gK} \right) \Phi(\phi,\Omega)
    \label{eq:mbs_trivial_spectrum}
\end{equation}
where $\textbf{K}$ and $\Omega$ are a wave vector and a cyclic frequency of a surface wave, $g$ is the acceleration due to gravity, $\phi$ is the azimuth of \textbf{K} and $\Phi$ is the angular spreading, normalized as follows: $\int_0^{2pi}\Phi(\phi)d\phi=2\pi$. The equation (\ref{eq:mbs_trivial_spectrum}) is implicitly or explicitly used in many papers on low frequency scattering since the surface waves PSD $S(\Omega)$ as a function of frequency is much more better studied than the 3D spectrum and linear dispersion law is asumed to be valid:
\begin{equation}
    \Omega^2 = gK
    \label{eq:mbs_linear_dispersion}
\end{equation}

To obtain some theoretical data on backscattering spectra, valid for the conditions of the experiment, the authors carried out a numerical simulation in the previous study \citep{Razumov2021}. The model was capable to account both for sound scattering in case of large surface displacements and for nonlinearity of surface waves. The exploited numerical model was 2D and the bottom reflection was not included.

\begin{figure}
	\centering
		\includegraphics[scale=0.8]{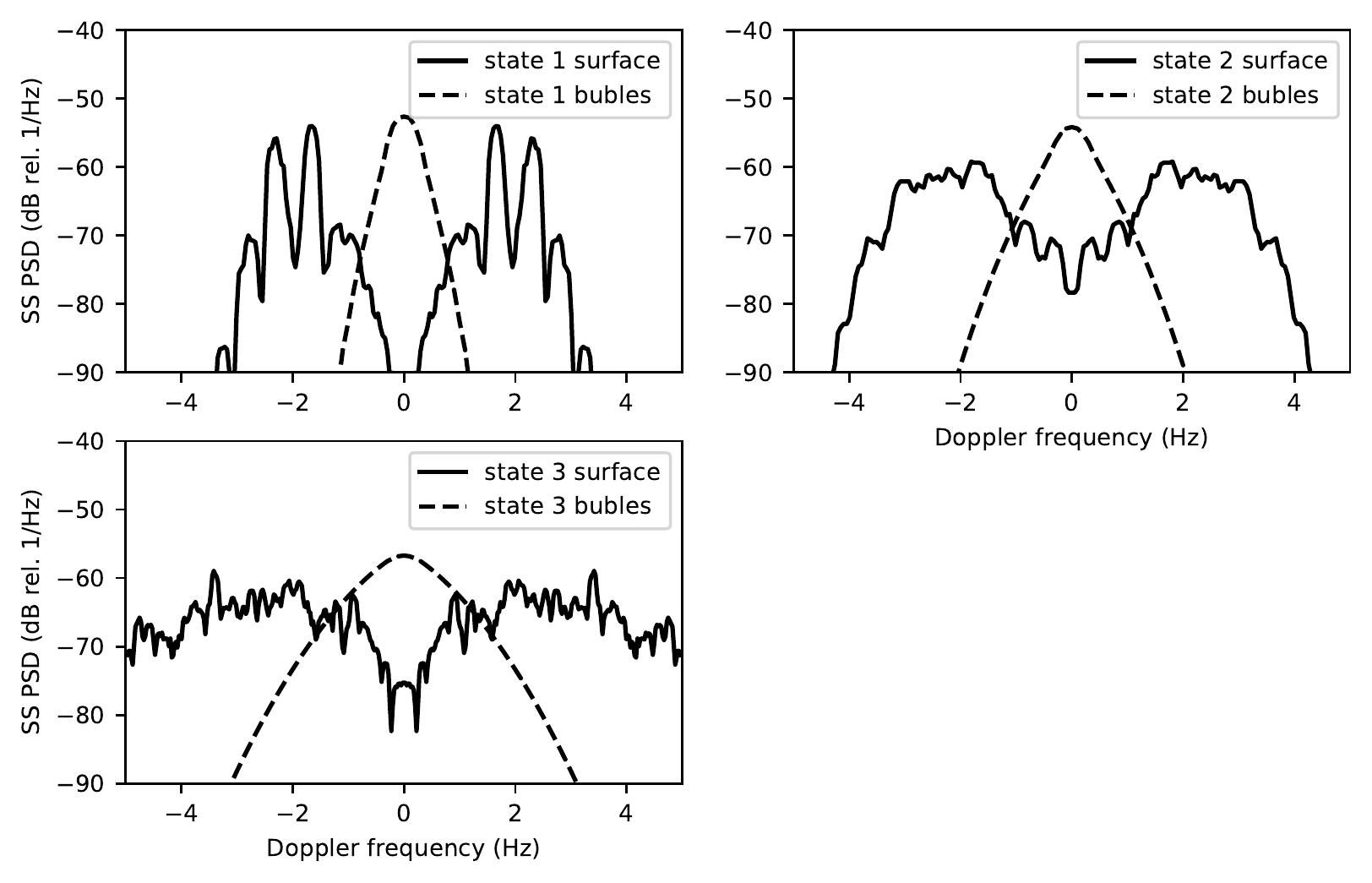}
	\caption{The numerical simulation results for backscaterring spectra, when surface roughness is determined by the Pierson-Moskowitz spectra with peak frequencies 0.6, 0.3 and 0.2 Hz (for states 1-3). The corresponding wind speeds are 2.28, 4.47 and 6.84 m/s; $P_r = 0.08$, 0.315 and 0.72. The sound frequency is 1.5 kHz, incident and scattering grazing angles are $30^\circ$. Two models are exploited, where scattering is originated by rough surface and bubble clouds. 
	}
	\label{fig:mbs_teor}
\end{figure}

The results, reproduced from \cite{Razumov2021}, are shown with solid lines on fig. \ref{fig:mbs_teor} for different sea states. (Dashed curves on the same figure are going to be discussed later.) Following \cite{Salin2012} the term of SS is extended to Doppler spreading of the scattered signal. Normalization is applied for the purpose of illustration due to there was a problem with the reference for absolute level in numerical simulation results. Integration will result in the value of SS = -55 dB for all curves plotted here.

The curve, labeled 'state 1 surface', shows approximately what we expect to obtain in terms of linear theory for sound scattering on the air-water interface, because $P_r = 0.08$ in this case. One can clearly see side lobes at Doppler frequencies of $\pm1.6$ Hz. These side lobes are expected due to the Bragg effect that together with the dispersion relation (\ref{eq:mbs_linear_dispersion}) leads to a common rule\footnote{Vertical angle of incidence is neglected as well}:
\begin{equation}
    f_{Bragg}=\frac{\sqrt{2gk}}{2\pi}
    \label{eq:mbs_f_Bragg}
\end{equation}
Some distortions of the spectrum has already appeared at this state and side lobes are widened and blurred in state 2 so that no maximum can be observed in state 3. The most significant effect is modulation \citep{Hwang2006}, since the resonant waves belong to the frequency band of 1.5-2 Hz and are shorter and lower by amplitude that the dominant ones. Moreover that, one can see from the plots that the center of mass for each lobe tends to move to higher frequency (by absolute value) with the wave height increase. The movement of each lobes center of mass appears to be more interesting effect than the widening of each lobe by itself.

The next effect that should be accounted is bubble scattering. First, they are moved by the mean water flow (mainly near-surface flow), that creates constant Doppler shift. This results in the shift of the Doppler centroid. Second, they are moved by oscillating field of currents, produced by surface waves. This creates double-side Doppler spreading with the dispersion proportional to the dispersion of the velocity:
\begin{equation}
v^2_{std} = \int \Omega^2 S(\Omega^2) d\Omega
\end{equation}
In most of cases $v_{std}$ increases with the increase of $H_{m0}$.

To obtain the shape of backscattering spectrum, created by subsurface bubble clouds, we use a computation routine, proposed in \cite{Salin2018}. The simulation results are plotted by dashed lines on fig. \ref{fig:mbs_teor}. Some parameters are not know for that model. No actual concentration of bubbles is used for the computation. Instead of that the resulting distribution is normalized to total level of SS = -55 dB, as above. Next, bubble concentration is assumed exponentially decaying with depth with e-times decay scale of 3 m, just as an example. As one can see from the plot, bubble oscillations due to surface waves currents are expected to create triangle-shaped spectra with not very high Doppler frequencies. As waves height increases, Doppler spreading increases too.

The resulting idea is that we actually have two concurrent models, which are scattering on surface waves, accounting for their nonlinearity, and scattering on subsurface bubbles that are driven by currents. We are not ready to state right now which one is the dominant one. (Mind that the difference in maximum levels of solid and dashed lines on fig. \ref{fig:mbs_teor} is not exact.) However both models predict the following behavior. Water flux results in the shift of the Doppler spectrum centroid. Increase in significant wave height results in increase of the spectral width. The listed effects are local we may use beamforming in the horizontal plane and time delay to make probe the surrounding water area.

%---------------------------------
\section{Materials}
The experiment took place on an oceanographic platform in the shelf zone of the northern part of the Black Sea in the second half of September 2021. 
The platform was located 600m from the shore (see fig. \ref{fig:map}). The depth near the platform is 30m, seaward for 2 nautical miles the depth of the shelf reaches 80m.
Sound speed profile, measured at 30 m depth, was uniform during the most of the time of the experiment. So the water was warm enough from the top to the bottom at this site due to it was the end of summer period. Starting form 27th of September a small portion of cold water started to appear in the bottom layer.

Contact measurements on the platform and remote sensing of the sea were carried out simultaneously by different types of sensors.
%The following quantities were measured on the platform: wind speed by a weather station, temperature in the water column by a vertical array of temperature sensors, water current by ADCP, and PSD of surface waves. 
%A string wave recorder and
A waverider buoy Datawell DWR-G4 located near the platform was used for studying surface waves. The following quantities were computed by the algorithm of the buoy 'out of the box':
\begin{enumerate}
    \item significant waveheight $Hm0$
    \item direction of the most intensive waves Dirp
    \item period of spectral peak $T_z$
\end{enumerate}
The buoy recorded these wave characteristics every 30 minutes, totaling about 400 measurements. 
Raw wave forms were recorded as well, but they are not used in this study directly.

\begin{figure}
	\centering
		\includegraphics[]{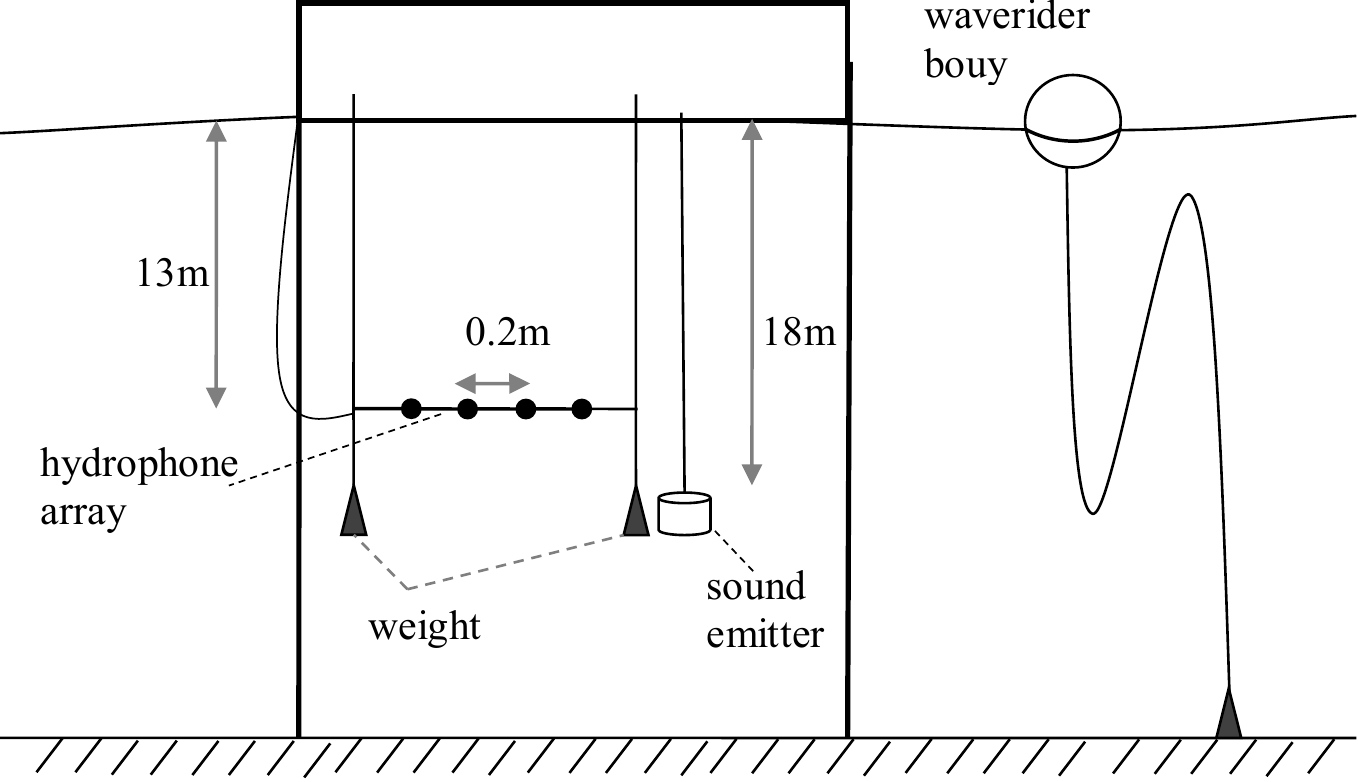}
	\caption{Experiment scheme in the vertical plane
	}
	\label{fig:scheme}
\end{figure}

\begin{figure}
	\centering
		\includegraphics[width=0.85\textwidth]{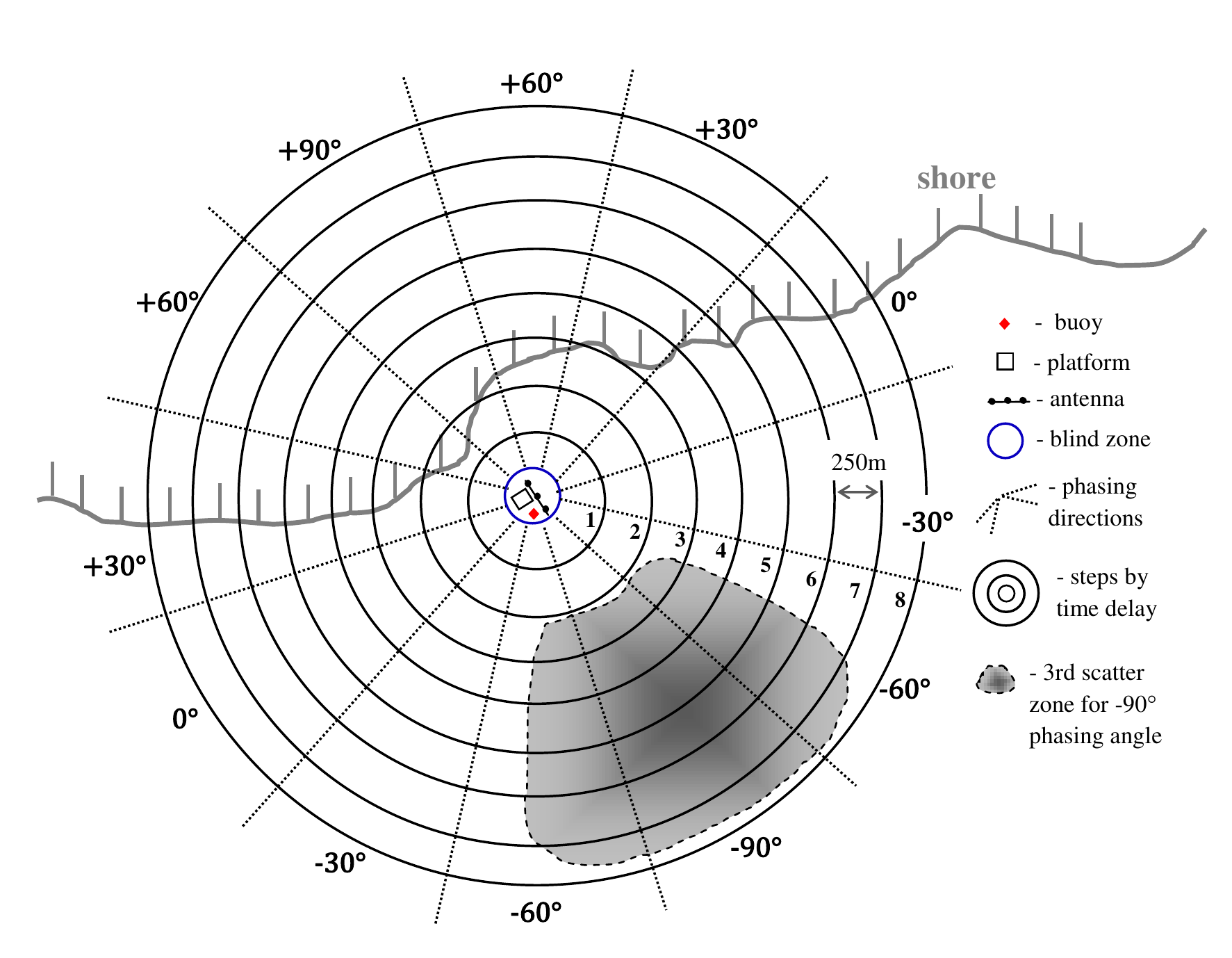}
	\caption{Experimental scheme, top view, phasing and time strobing grid. An example of approximate scattering region delineated for a pulse length of 2s
	}
	\label{fig:map}
\end{figure}

Underwater acoustic experiment was designed according to the classical monostatic scheme. 
The receiver was a horizontal array of hydrophones with a step of 0.2m, located at depth 13m (see fig. \ref{fig:scheme}). Effective number of exploited sensors was 13. Primal recording was done with a sampling rate of 24 kHz individually and synhronously for each hydrophone. Next complex envelopes with sampling rate of 50 Hz were computed. 

An omnidirectional ceramic ring-type sound emitter was located next to the receiver, 5m deeper. The produced acoustic pressure was around 1 kPa at 1 m. Output power was limited due to ecological reasons and since near located shore was a recreation area for local people. The transmitted signal was synthesized via sound card so it was possible to design the waveform to match the further described signal processing algorithm. A long tonal pulse mode was chosen. During each cycle of 90 s we transmitted 1320, 2020 and 2720 Hz CW signals per 2 s each and 2080 Hz per 8 s. One may see that the carrier frequencies of the sound corresponded to scattering by surface waves in the decimeter wavelength range.

The signals were transmitted in the same regime almost everyday and all day long (with short breaks for maintenance and conducting parallel acoustic measurements). Echoes were recorded correspondingly. So the acoustic data was collected in large amount and next we are going no discuss its relation to the data measured by the buoy. 

% --------------------------------
\section{Methods}

We apply machine learning to restore an unknown dependency function $\mathcal{F}$ between independent variables $X_i$ and depended variable $Y_i$, where $X_i$ is a 30 minutes long signal fragment at time moment $ i \in \mathcal{N} $, while $Y_i \in \mathcal{R} $ is an estimating value at time moment $i$. We assume $Y_i = \mathcal{F}(X_i) + e_i$, $e_i$ represents an additive error term. We logically split the function $\mathcal{F}$ by two parts: feature extraction process (sect. \ref{sec:feature_extraction}) and regression algorithm (sect. \ref{sec:regress}. Note that the feature extraction process can be considered as a function $\mathcal{F}': X_i \rightarrow X'_i $, where $X'_i \subset \mathcal{R}^{504}$ (504 corresponds to the total number of extracted features). 

We found that it is practical to redefine some of the buoy parameters for the use as an output of the regression model. Start with the squared wave frequency:
\begin{equation}
Y^{F2} = \frac{1}{T_z^2}
\label{wave_power}
\end{equation}
The idea is to treat $Y^{F2}$ as an effective wavenumber since it obeys the same power law as (\ref{eq:mbs_linear_dispersion}). The north projection of the effective wavenumber was calculated as 
\begin{equation}
Y^{\text{North} } = \cos(\text{Dirp}) \cdot Y^{F2}
\label{horizontal}
\end{equation}
While the East projection of the effective wavenumber was calculated as
\begin{equation}
Y^{\text{East}} = \sin(\text{Dirp}) \cdot Y^{F2} 
\label{vertical}
\end{equation}

We estimate each mentioned $Y$, e.g. component of the surface wave vector, separately by an individual regression model, while the feature extraction process we keep the same. Therefore we build and trained independently 4 regression models: to estimate North component $Y^{\text{North}}$, and East component of the surface wave vector $Y^{\text{East} }$, one to estimate the frequency of the waves $Y^{F2}$, and one for $\text{Hm}0$ parameter directly reported by buoy as well.

%-------------------------------------------
\subsection{Signal preprocessing and features extraction}
\label{sec:feature_extraction}

We rely on so-called pulse-Doppler signal processing. It means that the scattering intensity is measured using a proper band pass filter. Since the direct signal is intended to be narrow-band enough, it can be suppressed by selecting the Doppler frequencies of the interest. Scattering intensity is measured during the time interval with proper delay which is relative to a pulse transmission moment. In practice such processing is done vie short window Fourier transform and the spectrogram.

We start with the complex envelope with a sampling rate of 50 Hz at 4 carrier frequencies.
13  channels were used to form the directivity pattern in the directions from -90$^{\circ}$  to +90$^{\circ}$  with a step of 30$^{\circ}$ (The total number of channels was 32 but it was reduced due to technical problems.)
Then the spectrogram of the complex signal for each direction was calculated. The length of the window was 64 samples, the overlapping was 75\%. 
The moment of the beginning of each pulse was determined for time synchronization in order to cut and save each pulse.
Thus, at each carrier frequency the dataset is the power spectral density of the received audio signal $PSD(\tau, \Delta f, n, \theta )$, where $\tau$ is the delay, $\Delta f$ is the frequency shift relative to the carrier frequency, {\it n} is the pulse sequential number and $\theta$ is the phasing angle.

From section \ref{sec:theory} we have gained some ideas about the possible dependence of the shape of the backscattering spectrum on the state of the sea. The original dataset is transformed to obtain input data whose dimension comparable to the number of measurements.

\begin{figure}
	\centering
		\includegraphics[width=0.5\textwidth]{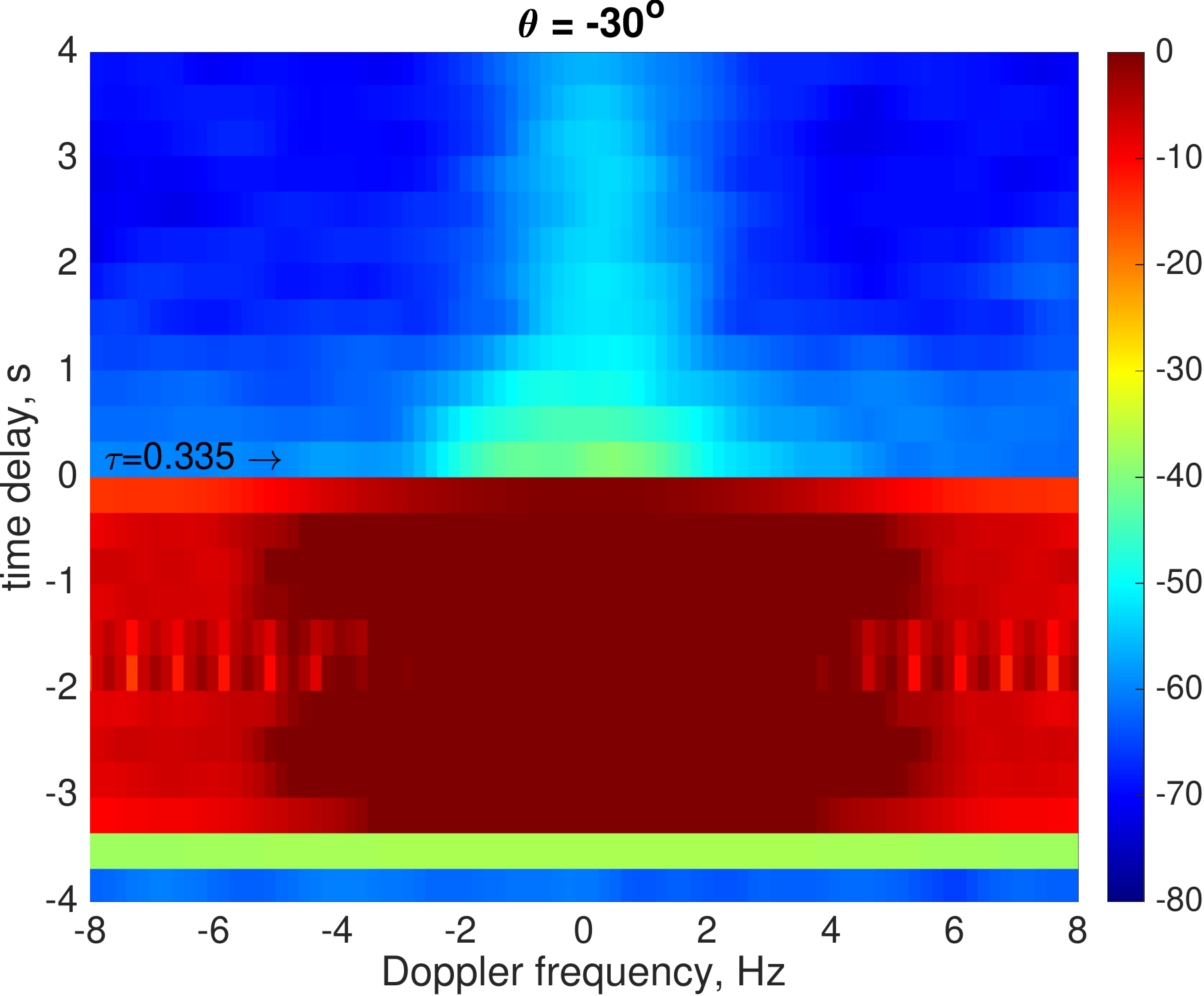}
	\caption{ An example of spectrogram of an acoustic signal
	}
	\label{fig:pulse}
\end{figure}

First, we average over about twenty pulses   $ \langle PSD(\tau,\Delta f, n, \theta )\rangle _{n} $ 
, which corresponds to a time step of  30 min. 
Bad pulses, which echos are distorted by noise, are excluded from averaging.
I.e. in the daytime, the broadband noise from motor boats, ships and the operations on the platform is especially noticeable compared to the signal. Such noise records are excluded based on the histogram criterion: the pulse is bad if the spectrogram has more than 5\%  of points exceeding the threshold value in the certain area on $\Delta f-\tau$ plane.  An example of the averaged reverberation spectrogram is shown on fig. \ref{fig:pulse}. The red zone, located on the plot from -3.3 to 0 over time axis, is the time period when the receiver is saturated by the initial signal. When we speak about such saturation, initial pulse length is expanded by the intervals, when a Fourier window overlaps the initial pulse. We analyze the spectrogram in the area of positive time values and until signal to noise ration allows us to do this.

Second, the dependence on the Doppler frequency $\Delta f$ is recalculated into spectrum features.
Scattering on bubbles leads to triangular-like shape of the spectrum, it is reasonable to describe it by a slope.
Further closer to the Bragg frequency, the slope may change due to the surface scattering. So we calculate the skew of the spectrum in the 2 frequency ranges.

The total level of backscattered signal and the signal level across the Bragg frequency can also be informative parameters. To estimate the central Doppler frequency, we introduce two frequency averages: the weighted average and the median average frequency. The last one should better take into account inequality or possible bumps on the tail of the distribution. 
The dependence on the angle $\theta$ and the time delay $\tau$ is preserved. 9 features for 7 directions and 8 time delays give us a total of 504 extracted features. 

\begin{figure}
	\centering
		\includegraphics[width=0.9\textwidth]{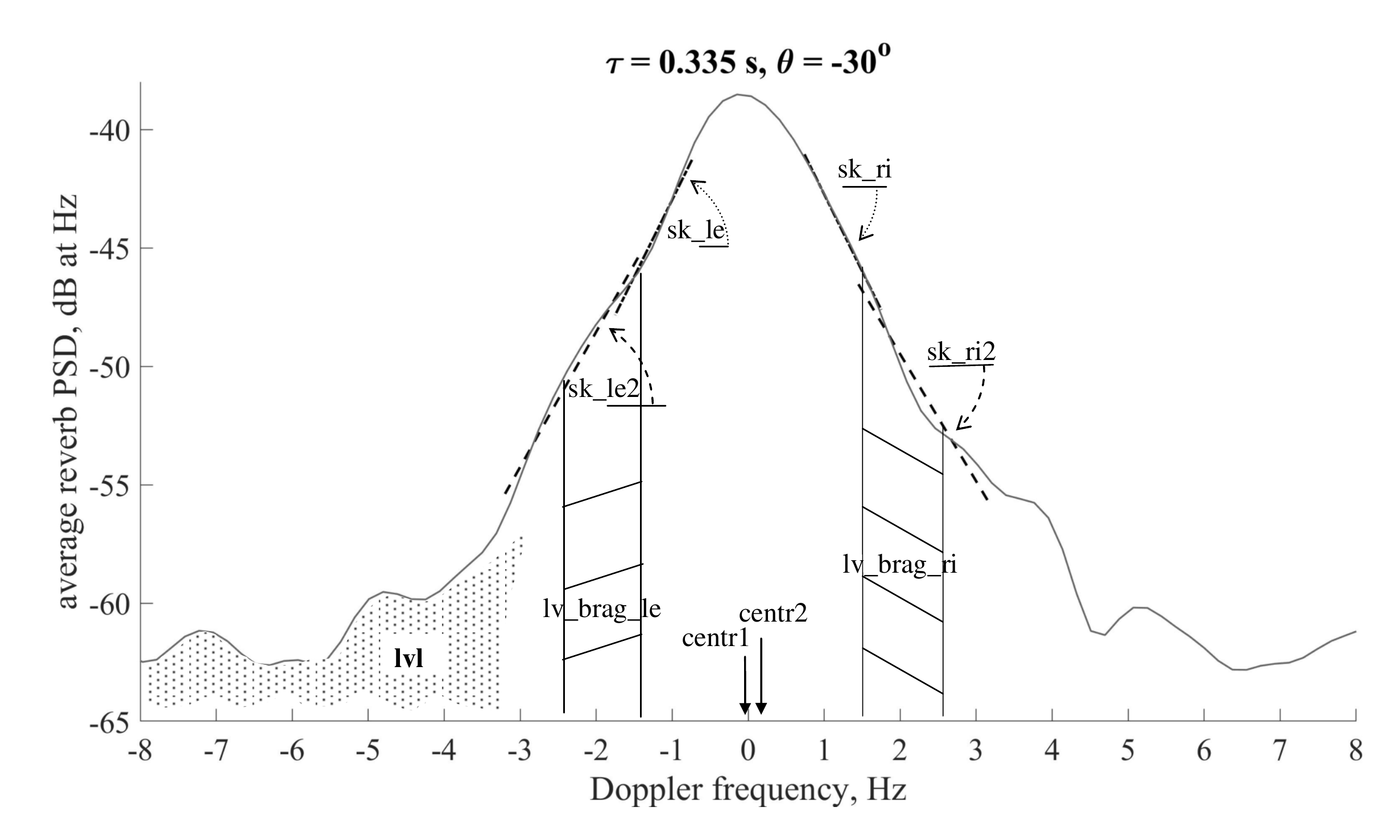}
	\caption{ An example of the scattering spectrum with a sketch, explaining how it can be described by the selected set of features
	}
	\label{fig:features}
\end{figure}

Figure \ref{fig:features} shows averaged by pulse PSD of reverberation in logarithmic scale taken from direction $-30^{\circ}$ with a time delay of 0.35s after pulse. Actually this plot is a section of a pattern like fig. \ref{fig:pulse}. Fig. \ref{fig:features} it explains how the features are computed.

We have accepted the following notation. Each feature is labeled as a literal string "name $\theta$ $\tau$. $\theta$ is given in degrees, $\tau$ in seconds from the end of the initial pulse. Name is one of the following:

1,2) \emph{sk\textunderscore le} \& \emph{sk\textunderscore ri} [dB/Hz] - spectrum skew ( the coefficient for the approximation of the polynomial of the first degree by the least squares method in Doppler frequency diapason [-1.75; -0.75] or [0.75; 1.75]) for negative and positive Doppler frequencies, respectively.  

3,4) \emph{sk\textunderscore le2} \& \emph{sk\textunderscore ri2} [dB/Hz] - the same features as 1,2) but calculated in a more distant frequency range ( [1.5; 3.2] for 2kHz carrier frequency)

5,6) \emph{centr1} \& \emph{centr2} [Hz] - weighted average and the median average frequencies, respectively.

7,8) \emph{lv\textunderscore brag \textunderscore le} \& \emph{lv \textunderscore brag \textunderscore ri} [dB] - signal levels in the range of 1Hz around the Bragg frequency ( -2±0.5 Hz \& 2±0.5 Hz respectively). It is worth mentioning that the illustration in the form of an area under the logarithmic PSD is inaccurate, in fact, the sum is calculated from PSD on a linear scale.  

9) \emph{lvl} [dB] - total level of backscattered signal.

However the horizontal line array, which is the available instrument, leads to the ambiguity in bearing estimation. Namely beamforming algorithm output is a function of the acoustic wave number projection on the line of the array. That leads to presence of two equal lobes in 0..360 degrees range. One can get only the sum of signals from these two directions. The directions, labeled as $-90^\circ$ and $90^\circ$ are free from such errors since they go along the line of the array, Yet antenna concentration is the worst in that directions.

%-------------------------------------------
\subsection{Regression model} \label{sec:regress}
% This subsection is for Alex.
The goal of regression analysis is to establish relationship between a dependent variable (outcome) and on or more independent variables (predictors, features). Regression model determines a functional space whereof the particular function is chosen by establishing a set of parameters. A linear regression is an example of the simplest case, but not the best one.
%wherever the functional space is a set of possible linear function $ y = \beta_0 + \beta_1 x_{1} + ... + \beta_p x_{p} + \epsilon$, defined by set of parameters $ \{ \beta_1, ..., \beta_p \} $.
In more complex case the function space can be non-linear. One of possible way to build non-linear function is to build a regression tree. Many algorithms to build regression tree that fit a function on observed data were developed. Most notable algorithms are ID3 \cite{quinlan1986induction}, C4.5 \cite{quinlan2014c4}, CART \cite{breiman2017classification}, and multivariate adaptive regression spline (MARS) \cite{friedman1991multivariate}. To use an ensemble of regression trees is a natural idea to increase accuracy. Extreme Gradient Boosting (XGBoost) \cite{chen2016xgboost} is one of the leading algorithm that based on this idea. It  has been used successfully for the last decades in different areas, including physical experiments \citep{roe2005boosted}. In the present paper we propose a method based on such model. We used a well known open source library \footnote{\url{https://github.com/dmlc/xgboost}} as an implementation of XGBoost algorithm. The linear regression was used as a baseline and it was implemented as a polynomial fitting. 

\begin{figure}
    \centering
    \includegraphics[scale=.40]{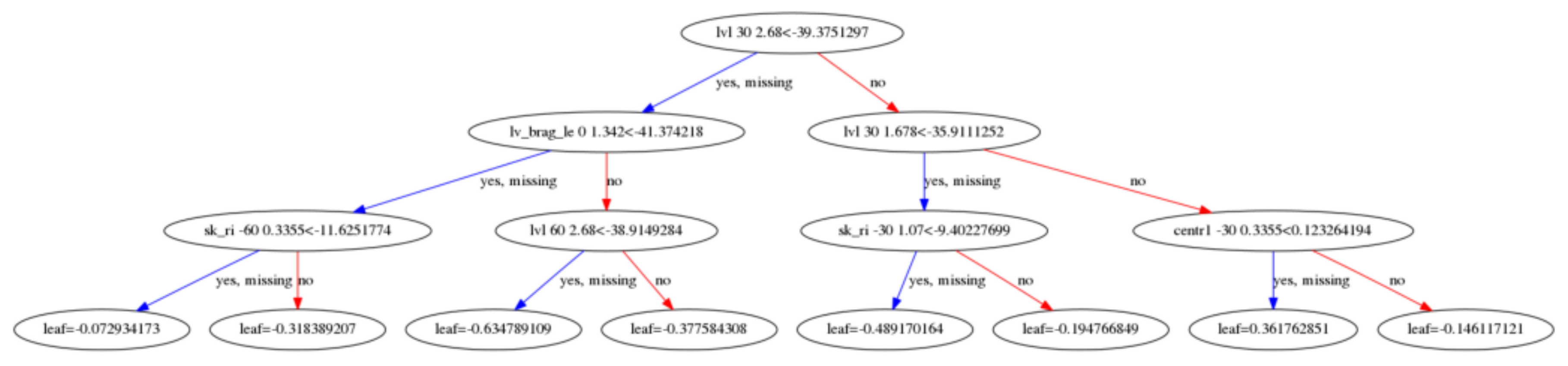}
    \caption{An example of the decision tree of depth 4 for North projection formed by XGBoost for 2020Hz frequency.}
    \label{fig:decision_tree}
\end{figure}

Figure \ref{fig:decision_tree} presents an example of a regression tree, built by XGBoost algorithm. The input data (namely the values of extracted features, listed above) is compared with some threshold values in multiple yes/no tests. The model should be trained, first. At this stage both input (acoustic features) and output (surface waves state) are given to the software. It selects which feature to compare in each cell and adjusts the threshold values in order to minimize a loss function (error function). Here, as a loss function we took the root mean square error (RMSE):
\begin{equation}
\label{eq:rmse}
RMSE = \sqrt{ \frac{1}{n} \sum^n_{i=1}(Y_i-\hat{Y_i})^2 }
\end{equation}

where $Y_i$ is observed value, while $ \hat{Y_i}$ is estimated value. The tree, plotted on fig. \ref{fig:decision_tree}, is a real example, produced by the software for our data. After training is done, the model is ready to make estimations. So, we feed the acoustic features as an input, and get the surface waves parameters as an output, which we compare with the real observed values.

%--------------------------------------------------------------------
\section{Results}

\subsection{Correlation analysis}
\begin{figure}[h]
	\centering
		\includegraphics[scale=0.8]{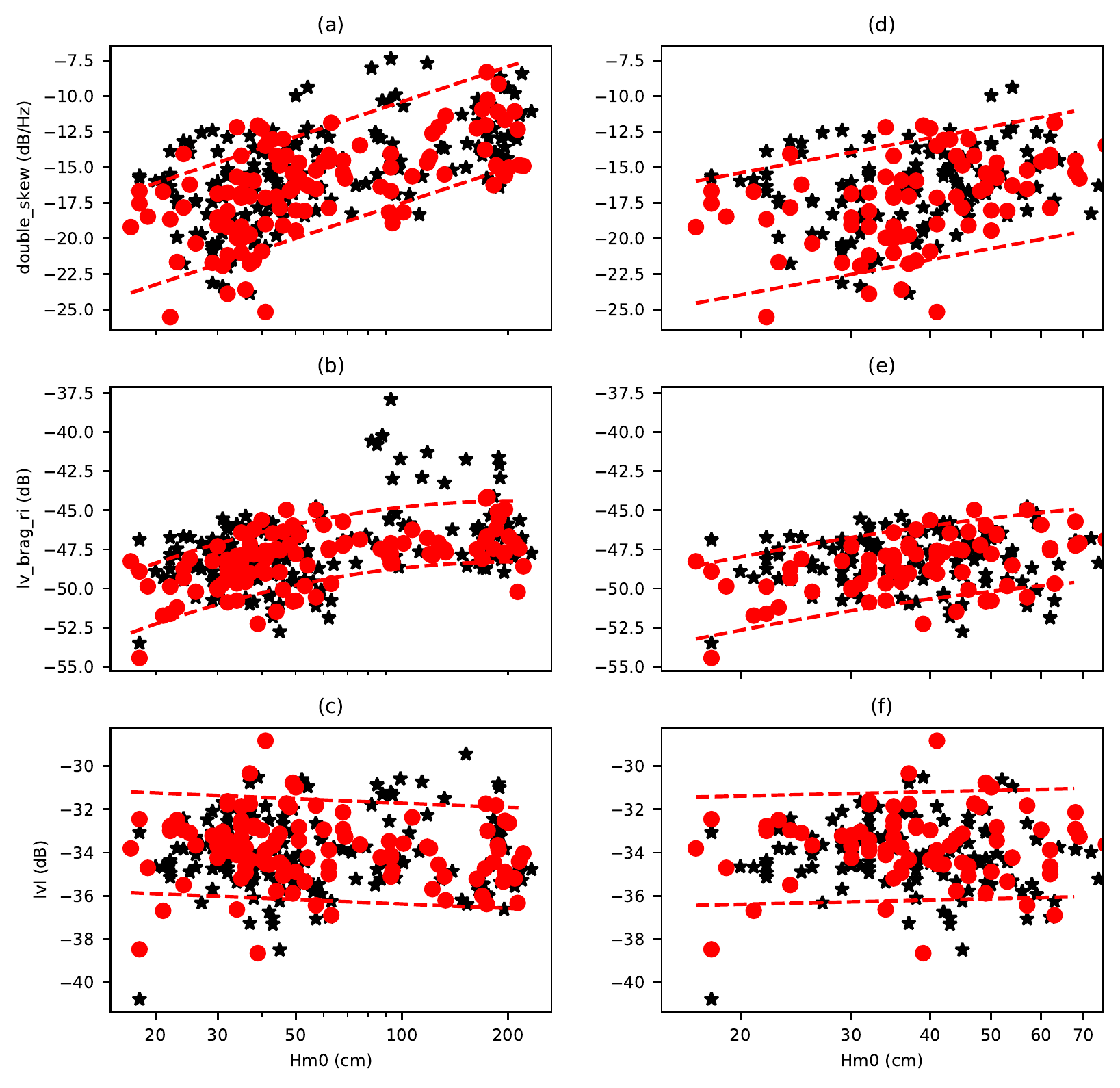}
	\caption{
    Dependence between features and a significant wave height. All features are obtained for the angle of $-60^\circ$ and the delay of 0.671 s. Red filled dots are the data without significant shift of Doppler centroid $(|centr1|<0.075 Hz)$ and stars represent the rest of the dataset. Dashed lines represent confidence intervals around a fitting line or curve. Right panels are scaled versions of left panels. }
	\label{fig:mbs_Hm0_corr}
\end{figure}

%\textbf{Mike}
Prior to exploiting machine learning methods we start from direct comparison of environmental parameters and acoustic signal features. 
Fig. \ref{fig:mbs_Hm0_corr} shows various features as functions of the significant wave height $H_{m0}$. Each marker on this and following plots corresponds to data, obtained in 30-min period during the whole experiment. 
The acoustic features, retrieved at smallest possible time delay, are chosen here to analyze backscattering form the area, that is close to the transducer and receiver, due to the instrument for direct measurement of $H_{m0}$ is located there.
Recall that negative angles are when scattering signal is received for the open sea, and at positive angles scattering takes place between the platform and the shore. 
 
As it has been said in section \ref{sec:theory}, spectral bandwidth of the scattered signal is expected to increase with the increase of the wave height. However,  due to the bottom scattering, which dominates on the zero Doppler frequency, a classical -3 dB bandwidth appeared to be not a very reliable parameter for the present dataset. Such bandwidth is not treated as a feature at all, and the other feature (fig. \ref{fig:mbs_Hm0_corr}a), describing how fast the spectral curve decays in 0.75-1.75 Hz span, is used instead of that one. Double skew is defined as $(sk\_ri - sk\_le)$. Its value is negative and approaches zero, when the spectrum becomes more gently sloping. Scattering level at the Bragg frequency is also examined (see fig. \ref{fig:mbs_Hm0_corr}b) since an increase in bandwidth should result in increase in side lobes either.

One may conclude that the spectral skew and the Bragg level are definitely correlated with the significant wave height. A linear or square parametrization is found for these features, using logarithmic scales on both axes. The plotted parametrization is found, using the dataset, obtained at mild sea states (namely the dataset plotted on the corresponding right panels: \ref{fig:mbs_Hm0_corr}d-f). Extrapolation of these data points match the other part of the dataset with over-meter wave heights. 90\% of points fall into the plotted confidence interval on the right column of plots, and 80\% on the left column of plots. 

\begin{figure}[h]
	\centering
		\includegraphics[scale=0.8]{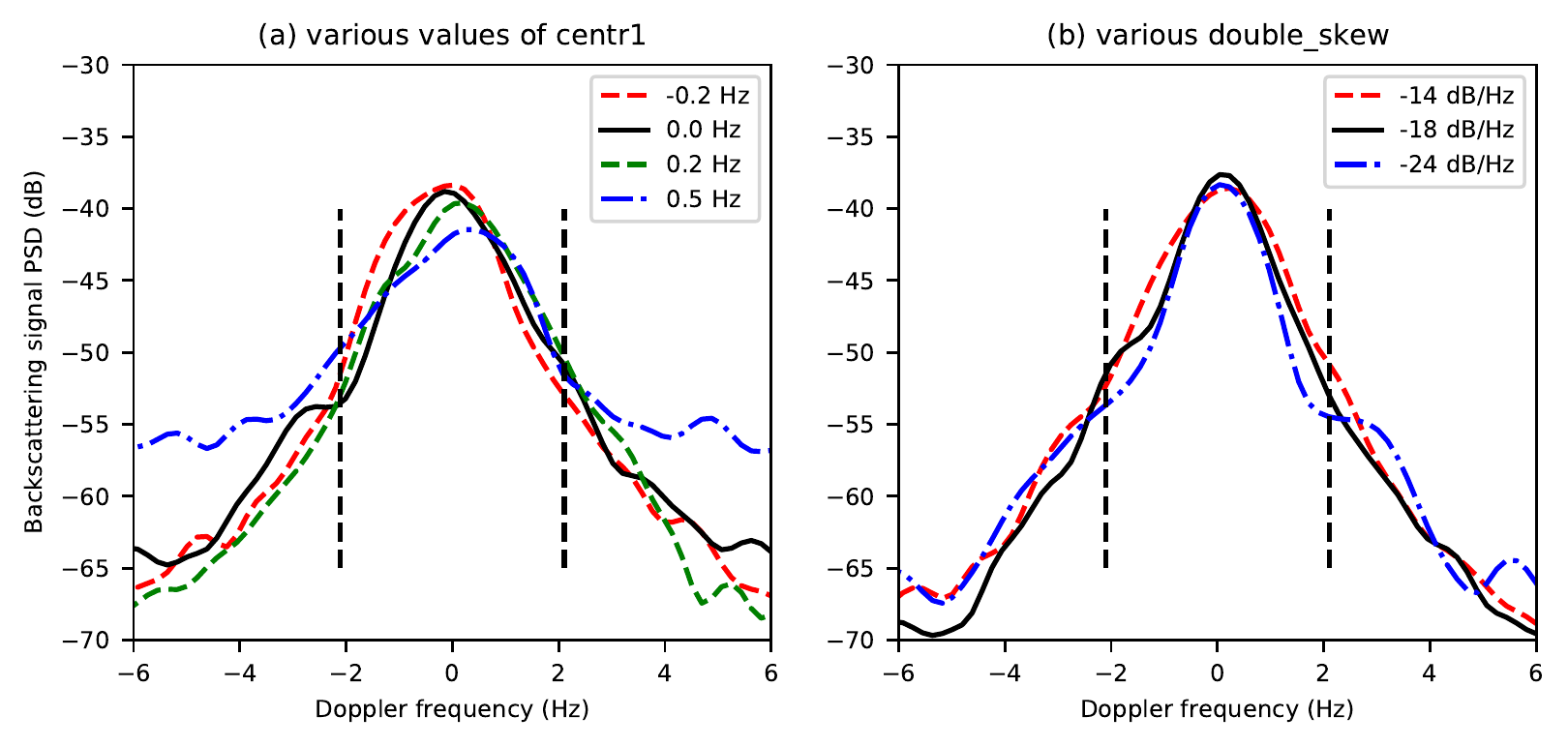}
	\caption{
    Examples of the backscattering spectra, obtained in the experiment (Pay attention to the feature values, printed in thelegend). Vertical dash lines point the Bragg frequencies for a reference. }
	\label{fig:mbs_extreme_and_mean}
\end{figure}

Exclusion of points with relatively large shift of the Doppler centroid helped to reduce the dispersion. The reason for that is the fact that the presence current results in change of all features. Most probable effect is that the whole  triangle-shaped spectrum is moved and more intensive scattering takes place of what we have expected to be the Bragg scattering. Surprisingly no correlation is found for total backscattering level (see fig. \ref{fig:mbs_Hm0_corr}c), despite many papers reports that this feature depends on the wind speed. Finally, examples of backscattering spectra, corresponding to some points on the above figures, are showed on fig. \ref{fig:mbs_extreme_and_mean}. Processing of these spectra resulted in extreme or{\it  vice versa} mean values of such features as Doppler centroid and spectral skew. 

From this section one may conclude that the backscattering spectrum turns out to be determined by a complicated mixture of physical effects. Attempt to find a direct link between single input and single output parameters, matched with the respect to the theory, demonstrated some correlation, but the precision is not enough for practical usage. Proper inversion (i.e. computing surface waves parameters from spectral features) can be done using the machine learning methods and this is going to be shown in the following subsection.

\subsection{Evaluation of regression models}

\begin{table}[]
\centering
    \caption{RMSE values for XGBoost regression equipped with 30 trees with max depth of 3 on cross validation with 5 folds}
    \label{tab:cross_val}  % The \label should go after the \caption in the float (table or figure)
    
    \begin{tabular}{lrrrr}
    \toprule
    Frequency & North projection & East projection & Squared wave frequency & Wave height \\
    (Hz) & $Y^{North}$ (Hz$^2$) & $Y^{East}$ (Hz$^2$) & $Y^{F2}$ (Hz$^2$) & $H_{m0}$ (cm) \\
    \midrule
    1320 &  0.0303 &  0.0595 &                0.0282 &    47 \\
    2020 &  0.0297 &  0.0612 &                0.0258 &    39 \\
    2080 &  0.0300 &  0.0585 &                0.0256 &    42 \\
    2720 &  0.0316 &  0.0580 &                0.0238 &    35 \\
    \bottomrule
    \end{tabular}
\end{table}

\begin{table}[]
\centering
    \caption{RMSPE (Root Mean Square Percentage Error)  values for XGBoost regression model equipped with 30 trees with max depth of 3 on the test set}
    \label{tab:rmspe}  % The \label should go after the \caption in the float (table or figure)
    
    \begin{tabular}{lrrrr}
    \toprule
    Frequency & North projection & East projection & Squared wave frequency & Wave height \\
    (HZ) & $Y^{North}$ (\%) & $Y^{East}$ (\%) & $Y^{F2}$ (\%) & $H_{m0}$ (\%) \\
    \midrule
    1320 &  288.3 &          78.0 &         \textbf{32.6} &    116 \\
    2020 &  332.6 &          76.1 &                36.8 &    \textbf{45.7} \\
    2080 &  233.4 &  \textbf{63.9} &                36.9 &    103 \\
    2720 &  \textbf{118.0} &  111.6 &                37.4 &    57.3 \\
    \bottomrule
    \end{tabular}
\end{table}

Accounting for several features together was achieved by feeding their values into the XGBoost algorithm. The model was build, trained and tested as it was described in the section \ref{sec:regress}. Common practice for such kind of testing is to split all experimental data into two non-overlapping parts, that are going to be used for training and for testing separately. Here the data was randomly split to train and test datasets by portion 10:1. 

Figures \ref{fig:compare_height_frequency} and \ref{fig:compare_height_frequency} show the results of testing stage. Various parameters of surface waves are plotted as they were estimated by model. Recall that the model is XGBoost preceded by the feature extractor, that processed the acoustic signal at available frequencies from 1320 to 2720 Hz (single frequency was tested at one time). The true values are plotted of the figures \ref{fig:compare_height_frequency} and \ref{fig:compare_height_frequency} as well. Only points, that belong to the testing dataset are plotted. There are gaps between them on the timeline, where the testing data was obtained. 

To evaluate the performance of the regression modes we used cross validation with 10 folds. Thus the data was spit by 5 no-overlapping parts. Each part was used as a test set for model evaluation, while 4 remaining parts where used as a training test to fit the model. For particular model we calculate the mean, and standard deviation over 5 RMSE values for each fold. RMSE values are presented in the table \ref{tab:cross_val}. 

In the table \ref{tab:rmspe} we present the values RMSPE  (Root Mean Square Percentage Error) that were calculated according to (\ref{eq:rmspe}) on the test set.
\begin{equation}
\label{eq:rmspe}
RMSPE = \sqrt{ \frac{1}{n} \sum^n_{i=1} \left(  \frac{Y_i-\hat{Y_i}}{Y_i}    \right)^2 }
\end{equation}
The most precise (see table \ref{tab:rmspe}) prediction was done for $Y^{F2}$, namely 33\% error was achieved which resulted in 16\% error of the determining the peak frequency. Next come significant wave height, estimated with a 46\% error. The worst result was for $Y^{North}$: the error is the same order of magnitude as the parameter value by itself.
This performance was obtained for the parameters: tree max depth of 4, number of estimators of 24, learning rate\footnotemark of 0.1.
\footnotetext{Learning rate influences to what extent newly acquired information overrides old information}

\begin{figure}
    \centering
    \includegraphics[scale=0.8]{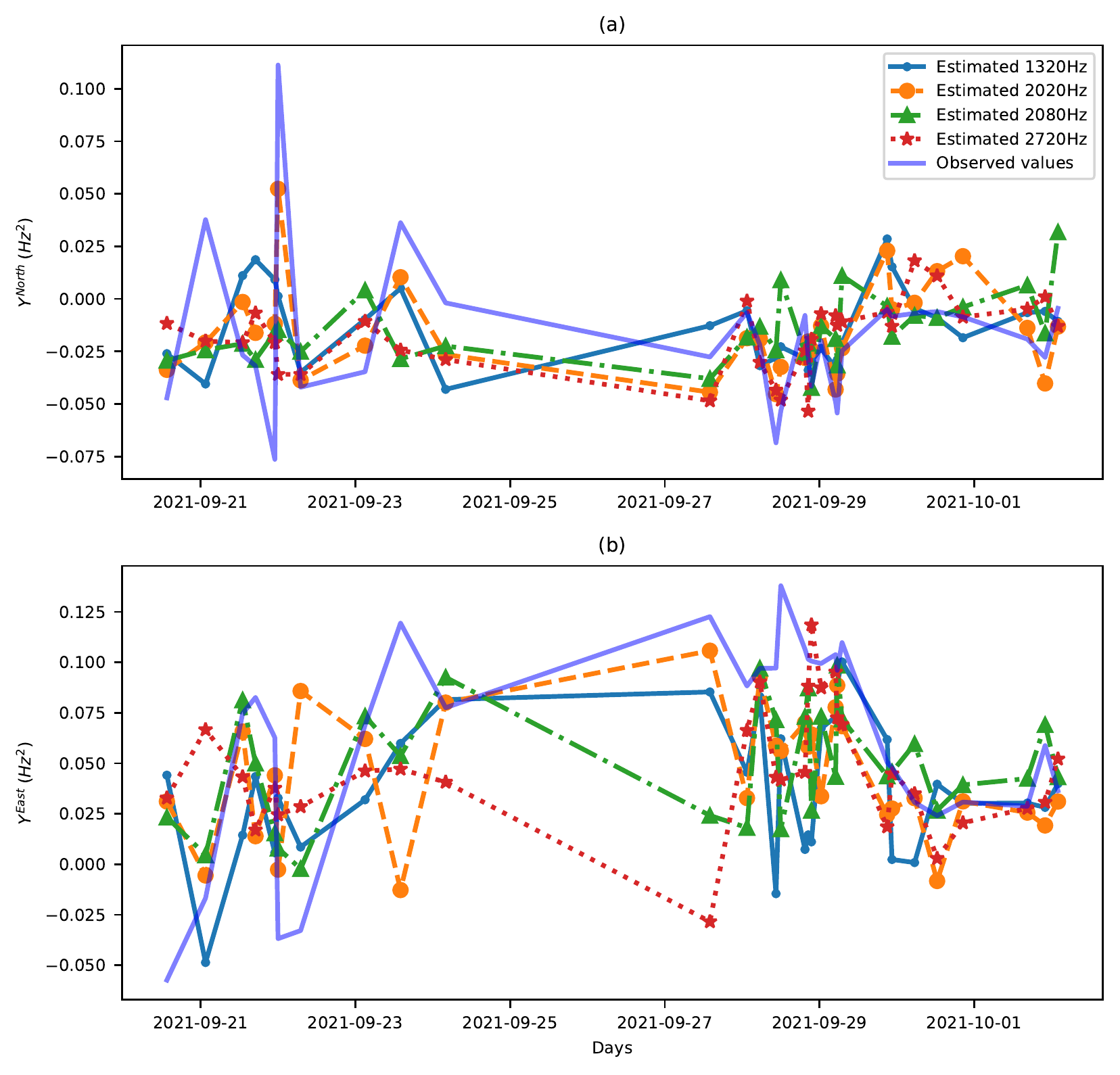}
    \caption{ Comparison of 4 XGBoost regression models equipped by 20 decision trees of the max depth 10 utilizing features for different frequencies to estimate the  projections of the effective wavenumber  }   
    \label{fig:compare NE_projections}
\end{figure}

\begin{figure}
    \centering
    \includegraphics[scale=0.8]{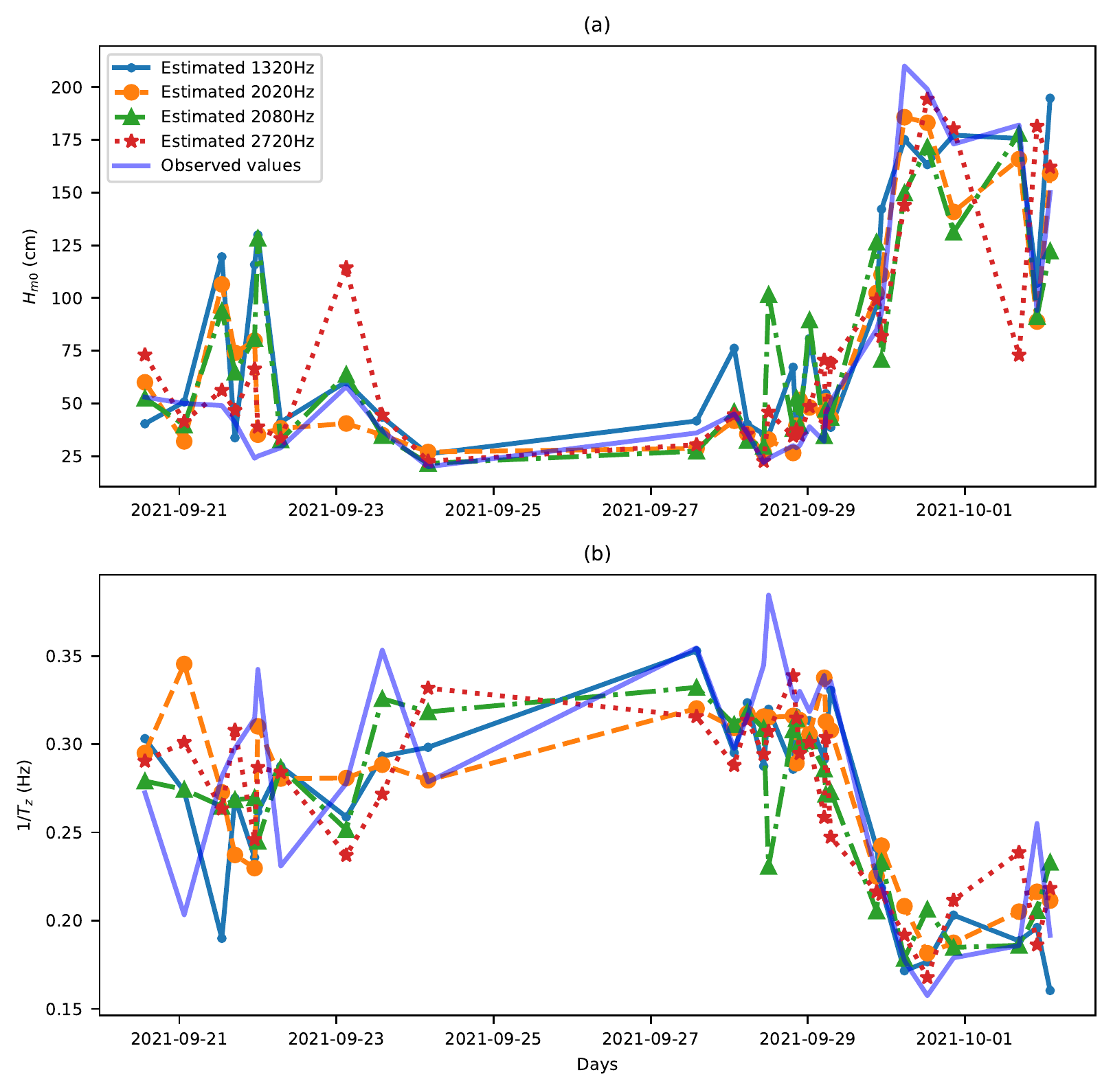}
    \caption{Comparison of 4 XGBoost regression models equipped by 20 decision trees of the max depth 10 utilizing features for different frequencies to estimate the values of Significant wave height and peak wave frequency.
    }
    \label{fig:compare_height_frequency}
\end{figure}

\begin{figure}[]
	\centering
		\includegraphics[scale=.50]{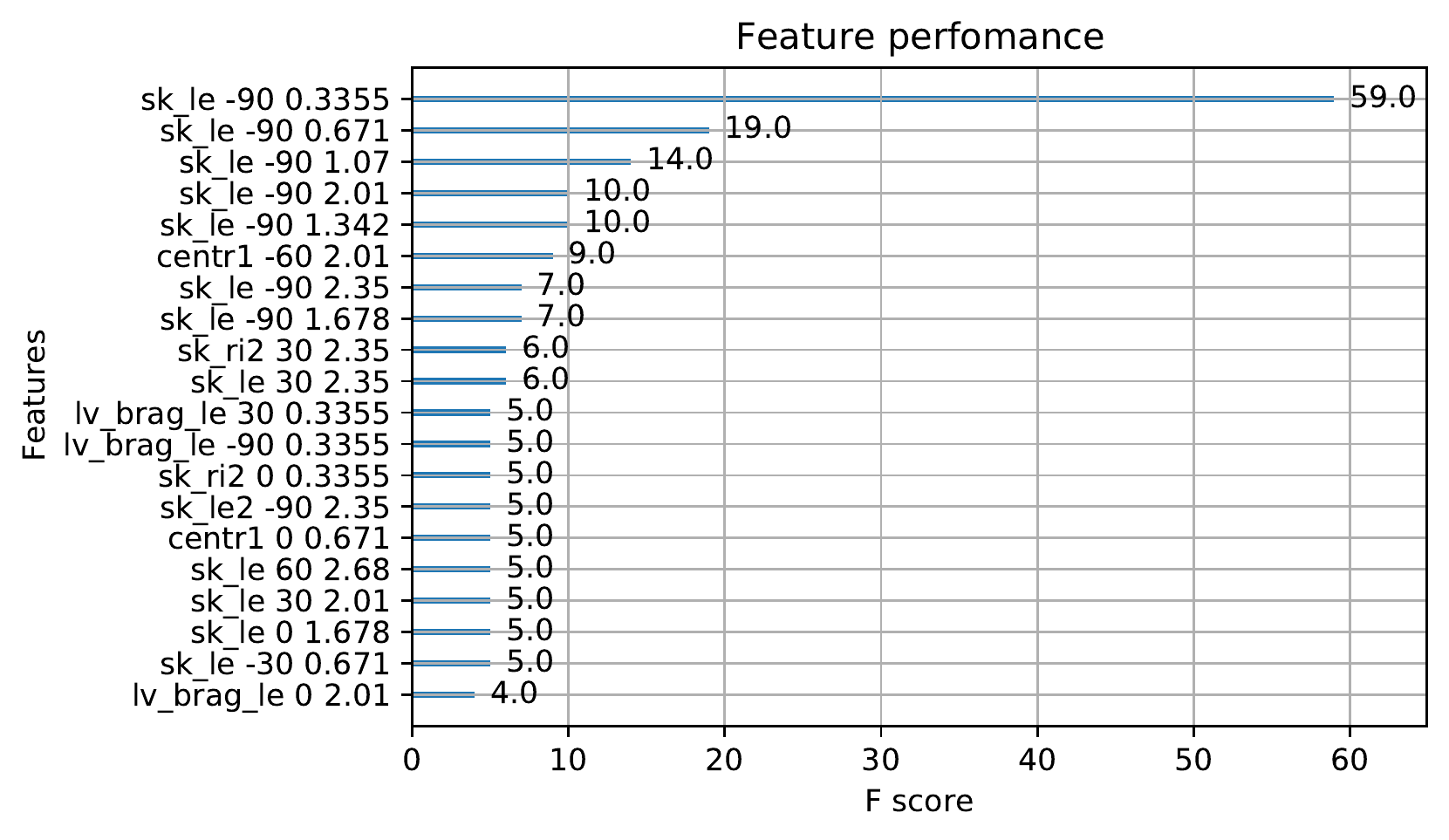}
	\caption{
	20 most important features reported by XGBoost algorithm. Frequency = 2720Hz, number of estimators = 20. The importance means how many times the feature was used in the decision trees to split the data.}
	\label{FIG:importance}
\end{figure}

\subsection{Parameters tuning}

\begin{table}[]
\centering
\caption{The results of parameters tuning of XGBoost regression model for predicting $H_{m0}$, done by grid search}  
\label{tab:tuning}
\begin{tabular}{llrrrrr}
\toprule
    &        &  mean RMSE & std of RMSE &  max\_depth &  n\_estimators &  learning\_rate \\
\midrule
%Hm0
    & 1320Hz &  40.4 &  37.3 &          3 &           193 &           0.10 \\
    & 2020Hz &  32.1 &  30.9 &          3 &           154 &           0.05 \\
    & 2080Hz &  38.6 &  32.3 &          3 &           276 &           0.05 \\
    & 2720Hz &  34.3 &  38.5 &          2 &           207 &           0.05 \\
%$Y^\text{North}$ & 1320Hz &   0.9509 &   0.8393 &          1 &            24 &           0.10 \\
%    & 2020Hz &   0.9500 &   0.8310 &          4 &            24 &           0.10 \\
%    & 2080Hz &   0.9497 &   0.8186 &          1 &            30 &           0.10 \\
%    & 2720Hz &   0.9540 &   0.8363 &          1 &            79 &           0.05 \\
%$Y^\text{East} $ & 1320Hz &   1.0028 &   0.8952 &          3 &             8 &           0.10 \\
%    & 2020Hz &   1.0392 &   0.8595 &          1 &             7 &           0.05 \\
%    & 2080Hz &   1.0007 &   0.8572 &          1 &             6 &           0.10 \\
%    & 2720Hz &   1.0413 &   0.8548 &          3 &             1 &           0.10 \\
\bottomrule
\end{tabular}
\end{table}

By using grid search with 10 foldes for cross validation we examined XGBoost regression model with the following parameters:
\begin{enumerate}
\itemsep=0pt
\item {Number of estimators} $ = 1, 2, ...,299$;
\item {Max depth of decision trees} $ = 1, 2, ...,10$;
\item {Learning rate} $ = 0.1, 0.05, 0.01 $
\item {Step size shrinkage $\eta$} $ = 0.3$
\end{enumerate}  

The table \ref{tab:tuning} presents results of the parameters tuning. The table shows that model is stable with the respect to its adjustable parameters.

\section{Discussion}
The described experiment showed principal possibility to estimate the parameters of energy-carrying surface waves using underwater sound of the mid-frequency band. We recognize that the we have provided comparison of local and remote measurements and that might be not completely correct.

The most remarkable fact is that the algorithm was not aware of existing physical models at all, nevertheless according to fig. \ref{FIG:importance} it had selected features, related to our prior theoretical expectations. E.g. spectral skew appeared to be a very useful feature for machine learning, as well as it had shown good results during the correlation analysis (fig. \ref{fig:mbs_Hm0_corr}). Moreover, features, corresponding to signals, coming from open sea and small delays were prioritized to estimate surface state in the center of coordinates. On the other hand, during the first attempt to process the data, we missed the time synchronization and the model did not produce any meaningful result until we fix that. So we do not overestimate the possibilities of the machine learning. It was the successful choice of features, based on the correct understanding of physical processes, that made it possible to train the model to make the correct prediction with such a small training sample. 

The measurement technique can be significantly improved. First, the design of the hydrophone array may be improved to avoid left-right ambiguity. E.g. one can use circular array or an array of vector sensors. Second, several transmitters and receivers are required for better measurement of both projections of the current speed in the same cell. Recall that the wavenumber projections of surface waves were estimated with low precision. Co-location of this measurement system with a coastal acoustic tomography system \citep{CAT2} may be suggested. Third, special frequency modulated signals may reduce the size of the measurement cell, however proper correspondence between Doppler scattering strength in case of tonal and frequency-modulated signals should be studied to do so \citep{Rui2016}. Careful filtering of bottom scattering should be provided as well.

After all, the underlying physics is still not completely clear. We started from two concurring models: surface Bragg scattering of sound and bubble scattering. We still cannot prove that one of them is dominant. If one have accepted the bubble scattering model, then one should expect the increase of scattering strength with at the conditions of strong wind and large wave heights due to increasing number of bubbles. However no significant increase of scattering strength was detect during all stages of the experiment \footnote{To be more precise, a 4 dB increase was found at some specific condition of strong wind with short fetch, but not enough statistical information was collected to make any conclusion from that event}. 

As for surface scattering, the Bragg resonance is that physical effect that should be mainly accounted in this frequency band. While studying the particular shape of Doppler backscattering spectra we have been able to see only moderate enhancement on Bragg frequencies, and that was observed only in cases of most calm sea states (see '-18 dB/Hz' and '-24 dB/Hz' curves of fig. \ref{fig:mbs_extreme_and_mean}b). 

Going dipper into to this detail we have found very few evidences of Doppler sidelobes in long-range {\it monostatic} reverberation in the available publications. \cite{Averbakh1990} probably published\footnotemark the best example of such spectrum with the Bragg peak. 
\footnotetext{The figures, originally printed by \cite{Averbakh1990}, were point-by-point reproduced by \cite{Salin2018} as parts of figs. 4 and 6, so the data is available for English-speaking readers.}
However there vast of side lobe evidences in long-range {\it forward} scattering experiments like conducted by \cite{Lebedev2004,Lynch2012}. Fine spectra are obtained in direct arrival conditions as well.

One more issue with our data is that a spectrum, received from one pulse in one resolution cell, appeared to be too stochastic, i.e. spectra varied greatly from pulse to pulse. Smooth curves were obtained only after averaging up to 20 pulses. However scattering took place on an area around a square kilometer, where the sound must have interacted with many scatterers, so we expected the result should have been averaged over an ensemble.

\section{Conclusion}
The described experiment showed principal possibility to estimate the parameters of energy-carrying surface waves using underwater sound of the mid-frequency band. In fact, what was done, is estimating the parameters of interest in the "center of coordinates" while processing acoustic reverberation spectra, received from remote points, located in about kilometer range from that point, due to existence of a blind zone. Machine learning method, called XGBoost, was exploited, and physical intuition helped to reduce the size of the input data. Training (or calibration) of a model was required prior to its usage for estimations (and at this stage the local data was used). 

The fact, how does the regression model behaves towards various input data, and the correlation analysis results as well encourages us for the further study. The signal processing algorithm is going to be reconfigured for estimating surface state in remote points. However one more experimental study with modified scheme is required to directly check the remote sensing function.

% <<<<<<<<<<<<<<<<<<<<<<<<<<<<<<<<<<<<<<<<<<<<<<<<<<<<<<<<<<<<<<<<<<<<<<<<<
\vspace{6pt} 

%%%%%%%%%%%%%%%%%%%%%%%%%%%%%%%%%%%%%%%%%%
%% optional
%\supplementary{The following are available online at \linksupplementary{s1}, Figure S1: title, Table S1: title, Video S1: title.}


%%%%%%%%%%%%%%%%%%%%%%%%%%%%%%%%%%%%%%%%%%
\authorcontributions{Supervision, A.E.; project administration, M.S.; investigation, A.E., D.K., D.R., M.S.; data curating, D.R.; software, A.P., D.R., formal analysis, A.P.; writing - original draft, A.P., D.R., M.S.}

\funding{This research was supported by the Russian Science Foundation, grant number 20-77-10081. 
The instruments for underwater acoustic measurements were provided to the team in terms of State Contract with the Ministry of Education and Science of the Russian Federation, grant number 0030-2021-0017.}

%\institutionalreview{for studies involving humans or animals.}
%\informedconsent{exclude this statement if the study did not involve humans.}

\dataavailability{See the repository \cite{our_data}} 

\acknowledgments{The authors are grateful to N.A. Bogatov, I.A Kapustin and A.A. Molkov for their help during the experiment and valuable discussions after it.}

\conflictsofinterest{The authors declare no conflict of interest.} 

%% Optional
%\sampleavailability{Samples of the compounds ... are available from the authors.}
%\abbreviations{Abbreviations}{  скопируйте снова образец из шаблона, если нужно добавить сокращения }

\begin{adjustwidth}{-\extralength}{0cm}

\reftitle{References}
\bibliography{paper1}

\vspace{12 pt}
{\it Originally based on the MDPI class for LaTeX files \href{mailto:latex@mdpi.com}{latex@mdpi.com} }

% If authors have biography, please use the format below
%\section*{Short Biography of Authors}
%\bio
%{\raisebox{-0.35cm}{\includegraphics[width=3.5cm,height=5.3cm,clip,keepaspectratio]{Definitions/author1.pdf}}}
%{\textbf{Firstname Lastname} Biography of first author}
%
%\bio
%{\raisebox{-0.35cm}{\includegraphics[width=3.5cm,height=5.3cm,clip,keepaspectratio]{Definitions/author2.jpg}}}
%{\textbf{Firstname Lastname} Biography of second author}

% For the MDPI journals use author-date citation, please follow the formatting guidelines on http://www.mdpi.com/authors/references
% To cite two works by the same author: \citeauthor{ref-journal-1a} (\citeyear{ref-journal-1a}, \citeyear{ref-journal-1b}). This produces: Whittaker (1967, 1975)
% To cite two works by the same author with specific pages: \citeauthor{ref-journal-3a} (\citeyear{ref-journal-3a}, p. 328; \citeyear{ref-journal-3b}, p.475). This produces: Wong (1999, p. 328; 2000, p. 475)

\end{adjustwidth}
\end{document}